\documentclass[a4paper]{jpconf}
\usepackage{graphicx}
\usepackage{txfonts}
\usepackage{epstopdf}
\usepackage{epsfig}
\usepackage{subfigure}
\usepackage{array}

\begin{document}

\title{[FeII] as a tracer supernova rate}

\author{M~J~F Rosenberg, P~P van der Werf and F~P Israel}

\address{Sterrewacht Leiden, Universiteit Leiden, P.O. Box 9513, NL-2300 RA Leiden, The Netherlands}

\ead{rosenberg@strw.leidenuniv.nl}

\begin{abstract}
Supernovae play an integral role in the feedback of processed material into the ISM of galaxies and are responsible for most of the chemical enrichment of the universe.  The rate of supernovae can also reveal the star formation histories.  Supernova rate is usually measured through the non-thermal radio continuum luminosity, but in this paper we establish a quantitative relationship between the [FeII]$_{1.26}$ luminosity and supernova rate in a sample of 11 near-by starburst galaxies.  SINFONI data cubes are used to perform a pixel pixel analysis of this correlation. Using Br$\gamma$ equivalent width and luminosity as the only observational inputs into Starburst 99, the supernova rate is derived at each pixel and a map of supernova rate is created. This is then compared morphologically and quantitatively to [FeII]$_{1.26}$ luminosity map.  We find a strong linear and morphological correlation between supernova rate and [FeII]$_{1.26}$ on a pixel-pixel basis:
 \[ log\frac{\nu_{SNrate}}{yr^{-1}pc^{-2}} = \left (1.01 \pm 0.2\right )\ast  log\frac{[FeII]_{1.26}}{erg s^{-1}pc^{-2}} - 41.17 \pm 0.9\] 
The Starburst 99 derived supernova rates are also in good agreement with the radio derived supernova rates, which further demonstrates the strength of [FeII] as a tracer of supernova rate. With the strong correlation found in this sample of galaxies, we now qualitatively use [FeII]$_{1.26}$ to derive supernova rate on either a pixel-pixel or integrated galactic basis.
\end{abstract}

\section{Introduction}

Supernova rate (SNrate) is typically estimated by the integrated non-thermal radio continuum emission.  The connection between radio continuum emission and SNrate comes from the infrared-radio relation that shows a tight correlation between star formation rate and both radio and infrared emission \cite{1992ARA&A..30..575C}, and references therein). In addition, near-infrared (NIR) observations of supernova remnants (SNR) often show strong [FeII] emission line flux coincident with the radio peak.  Iron atoms are typically locked in dust grains and shock fronts of the SNR cause efficient grain destruction through thermal sputtering.  This releases the iron into the gas-phase where it is singly ionized by the interstellar radiation field, making [FeII] a strong tracer of shocks, which are often caused by expanding SNR.  Imaging and spectroscopy point to [FeII] emission being coincident with known radio SNR, indicating that [FeII] can provide at least a relative measure of SNrate.  

\section{Sample}
\label{sec:sample}
We have selected 11 bright near-by starburst galaxies at distances ranging from ~10-100 Mpc.  A summary of the galaxies included in this study can be found in Table~\ref{table:galinfo}.  All galaxies are spiral galaxies ranging from S0-a to Sd with the exception of NGC 520, which is a merger galaxy.  There are two edge-on galaxies in our sample, NGC 520 and NGC 3628.  The far-infrared luminosities range from $1.4 \times10^{10} L_\odot$ to $3.2\times10^{12} L_\odot$, where Arp 220 is considered an Ultra Luminous Infrared Galaxy (ULIRG), with $L>10^{12} L_{\odot}$, and NGC 6240, NGC 1614 and NGC 7552 are classified as Luminous Infrared Galaxies (LIRGs), with $L>10^{11} L_{\odot}$,.  Our sample excludes galaxies with very active nuclei, however we have four Low-Ionization Nuclear Emission-line Region (LINER) galaxies and one weak Seyfert 2 (Sy2) nucleus.
\begin{table*}
 \caption{Summary of galaxy parameters.}
 \begin{center}
\begin{tabular}{llllllll}
\br
Galaxy   &  Morph.& Activity &Dist.  & \textit{i} &log L$_{FIR}$ \\
        & J2000 & & [Mpc] & [$^\circ$] & [L$_\odot$]\\
\mr

NGC 3628 &Sb pec edge-on&HII LINER&12.8&847$\pm$2&79.29&10.14\\
NGC 4536 &SAB(rs)bc&HII Sbrst&15.4 &1802$\pm$3&58.9&10.17\\
NGC 1792 &SA(rs)bc&HII&13.2&1210$\pm$5&62.78&10.22\\ 
NGC 1084 &SA(s)c&HII&16.6&1409$\pm$4&46&10.42\\ 
NGC 1808 &(R)SAB(s)a&HII&12.3&1001$\pm$4&83.87&10.55\\ 
NGC 520  &pec&Merger Sbrst&30.5&2162$\pm$4&77.49&10.81\\
NGC 7552 &(R')SB(s)ab&HII LINER&22.5&1611$\pm$6&23.65&11.03\\
NGC 7632 &(R')SB(s)0$^0$&HII &19.3&1535$\pm$15&82.44&11.43\\
NGC 1614 &SB(s)c pec&Sbrst&64.2&4778$\pm$6&41.79&11.43\\
NGC 6240 &S0-a&LINER&108.8&7242$\pm$45 &73.0&11.73\\
Arp 220  &Sd& LINER Sy2&82.9&  5420$\pm$6& 57&12.50\\    
\br
\end{tabular}
\end{center}
\label{table:galinfo}
\end{table*}

\section{Methods}
\label{sec:methods}
 \subsection{Extinction Correction}

In order to study the true emission line strengths, it is crucial to determine the amount of dust extinction obscuring each nucleus.  The extinction is calculated at each pixel and an extinction map is created showing the regions most effected by dust.  The A$_V$ is determined using the extinction law at NIR wavelengths of $A_{\lambda}\propto\lambda^{-1.8}$ \cite{1990ApJ...357..113M}.  

To grasp the true morphology and magnitude of the line emission, extinction-corrected [FeII]$_{1.26}$ and Br$\gamma$ linemaps are created using the Pa$\beta$/Br$\gamma$ line ratio.  The [FeII] map is used directly to find the [FeII] luminosity at each pixel.  The Br$\gamma$ linemap, on the other hand, is used along with the K band continuum to calculate both the Br$\gamma$ equivalent width and the Br$\gamma$ luminosity, which are used as the \emph{only} observational inputs to calculate SNrate from the Starburst 99 model, described in detail in Section~\ref{sec:sb99} below.   

\subsection{Calculating SNrate}
\label{sec:sb99}
Starburst 99 (from here on referred to as SB99) is a tool that models spectrophotometric properties of star forming galaxies \cite{sb99} such as spectral energy distributions (SEDs), luminosities, equivalent widths, supernova rates and colors.  It includes predictions of the variations in these properties as a function of starburst age.  These models are calculated for two extreme star formation modes: continuous, where star formation proceeds continuously at a constant rate, and instantaneous (delta function) starburst.

A near solar metallicity ($Z=0.02$), Salpeter IMF ($\alpha=2.35, M_{low}=1 M_{\odot}, M_{up}=100 M_{\odot}$) and an instantaneous star formation mode are used.  For the case of an instantaneous starburst, the burst is normalized to an initial starburst mass of $10^6$ M$_{\odot}$.   Using the observed Br$\gamma$ equivalent width (EW(Br$\gamma$)), which is independent of this normalization, the average age of the population dominating the emission in each pixel is calculated. From age, SB99 then provides the expected SNrate. Because of the normalization, the SNrate must be appropriately scaled by comparing the SB99 age-dependent prediction of ionizing photon flux (N(H$^{\circ}$)) to the observed ionizing photon flux, which scales with Br$\gamma$ luminosity.  This comparison allows a scaling factor that is directly proportional to the initial mass and initial star formation rate, allowing the SNrate to be scaled based on the true initial conditions of the region represented in each pixel. 

\section{Analysis}
\label{sec:analysis}
\subsection{Age and SNrate}
Table~\ref{tab:snrate} presents the EW(Br$\gamma$) followed by the SB99 results, including the average age and integrated SNrate using the instantaneous burst model for each galaxy.  Although the analysis was done on a pixel-pixel basis, the numbers presented in Table~\ref{tab:snrate} are averaged over the galaxy, in the case of age and equivalent width, and integrated over the galaxy in the case of SNrate. 

\begin{table}
\centering
\caption{Average equivalent width of Br$\gamma$ and the SB99 derived average ages and integrated SNrates using the instantaneous starburst model.}
 \begin{tabular}{|l||c|c|c|}
\hline
Galaxy  &  EW(Br$\gamma$)  &    Age$_{inst}$   & SNrate$_{inst}$  \\ 
        &      \AA{}       &   (Myr)        &   yr$^{-1}$     \\
\hline
NGC 7632& 5.9             &   7.0           &    0.03        \\
NGC 3628& 7.5             &   6.9           &    0.01         \\
NGC 4536& 10.6            &   6.7           &    0.3        \\
NGC 1792& 4.1             &   7.9           &    0.007         \\
NGC 1084& 4.8             &   7.3           &    0.009       \\
NGC 1808& 7.2             &   6.9           &    0.06          \\
NGC 520 & 11.7            &   6.6           &    0.3       \\
NGC 7552& 13.7            &   6.7           &    0.3         \\
NGC 7632& 5.9             &   7.0           &    0.03        \\
NGC 1614& 22.9            &   6.4           &    0.9          \\
NGC 6240& 3.6             &   7.7           &    3.6         \\
Arp 220 & 8.0             &   6.8           &    0.7       \\
\hline
 \end{tabular}
\label{tab:snrate}

\end{table}

\subsection{Qualitative Correlation}

\begin{figure*}
\centering
\includegraphics[width=17cm]{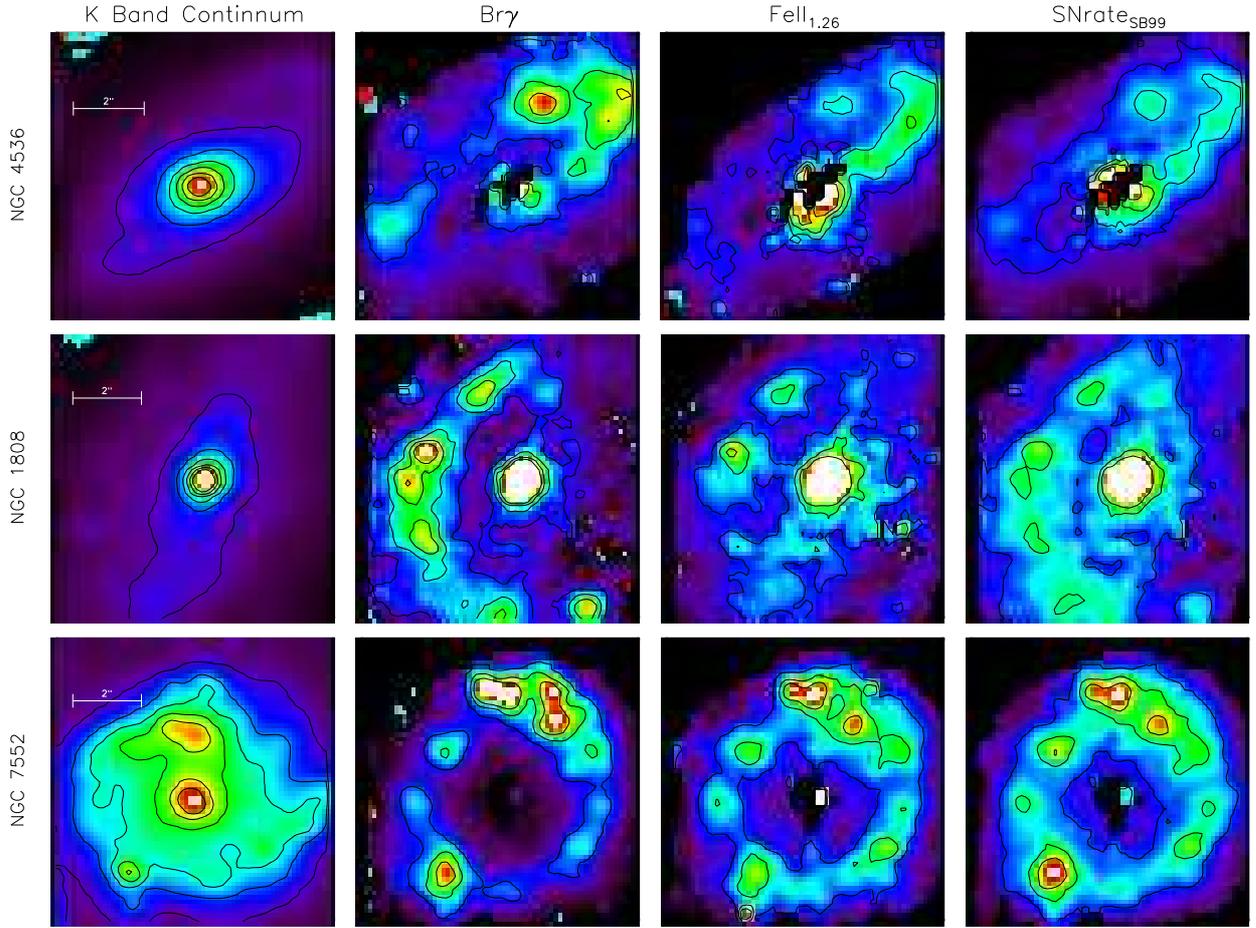}
\caption{A comparison between the K-band continuum, Br$\gamma$ linemap, [FeII]$_{1.26}$ linemap, and the SNrate morphologies for NGC 4536, NGC 1808, and NGC 7552.}
\label{fig:snr}
\end{figure*}

To demonstrate the morphological correlation between the [FeII] and SNrate, a side-by-side comparison of the K band continuum, extinction corrected Br$\gamma$ flux, extinction corrected [FeII]$_{1.26}$ flux, and SNrate is shown in Figure~\ref{fig:snr} for NGC 4536, NGC 1808, and NGC 7552.  These galaxies were selected for presentation because of the high quality of the observations.  Visually comparing the morphologies of the four different maps, we find that the SNrate most closely resembles the [FeII] emission.  

Focusing on NGC 4536, the Br$\gamma$ emission has a bright knot directly north of the nucleus with secondary emission peaks to the north-west and south-east of the nucleus.  The [FeII] emission is concentrated around the galactic center with a long plume of emission coming from the nucleus and extending towards the north-west corner of the galaxy.  There is a small knot of faint [FeII] emission at the location of the Br$\gamma$ peak.  Similar to the [FeII] emission, the SNrate peaks around the center with an arm extending towards the north-east corner.  

NGC 1808 displays a clear starburst ring in Br$\gamma$ surrounding the bright nucleus.  The [FeII] reveals a much less prominent ring configuration with more extended emission around the nucleus spanning the south-east to north-west plane.  The SNrate map shows more ring structure than the [FeII] but it is quite diffuse and lacks the strength of the Br$\gamma$ peak in the north-east region of the ring.  

The face-on spiral, NGC 7552, also reveals a starburst ring, with little Br$\gamma$ emission in the nucleus.  Focusing on the northern Br$\gamma$ knots, we see three 3 main knots, two slightly north-west of the nucleus, and one elongated knot directly north of the nucleus.  Comparing the Br$\gamma$ emission to the [FeII] emission, we discuss only one of the two north-western Br$\gamma$ knots.  In addition, the morphology in the elongated knot differs from that of [FeII] and appears as two distinct knots.  As shown by the SNrate morphology, the northern emission knots resemble the realtive flux and morphologies of the [FeII] emission much more closely than the Br$\gamma$ emission.  Specifically, the top north-western knot is also missing in the SNrate map and the elongated northern Br$\gamma$ peak is resolved into two peaks, resembling the relative flux ratios seen in [FeII].  

The comparison of the Br$\gamma$ and [FeII] emission to the morphology of the SNrate per pixel reveals a general similarity between the [FeII] emission and SNrate, a much closer correlation than the [FeII] and Br$\gamma$ or the SNrate and Br$\gamma$. We emphasize again the \emph{only} observational input into the SNrate calculation is the Br$\gamma$ luminosity and EW(Br$\gamma$), the [FeII] flux is never used. However, the SNrate correlates more with the [FeII] emission than the Br$\gamma$.  This strongly supports [FeII] as a robust tracer of SNrate.      

\subsection{Quantitative Correlation}

To test if the [FeII] to SNrate correlation holds quantitatively, we compare the [FeII]$_{1.26}$ luminosity to the SNrate, pixel by pixel.  The linemaps were not additionally filtered with the exception of NGC 6240, which was filtered in regions where the Br$\gamma$ flux is insufficient to determine the true extinction.  The pixel-pixel plot of SNrate against [FeII] luminosity for each galaxy is shown in Figure~\ref{fig:pixpix}.  Both the SNrate and [FeII] luminosity are scaled per square parsec such that each pixel represents the same physical area.  A least squares regression was performed in log space combining the points from all the galaxies excluding NGC 1792, due to the poor signal-to-noise linemaps.  This combined linear regression is plotted by a solid black line in Figure~\ref{fig:pixpix} for direct comparison and the relation is shown below.  

\begin{equation}
log\frac{\nu_{SNrate}}{yr^{-1}pc^{-2}}=\left ( 1.01 \pm 0.2 \right ) \ast
log\frac{[FeII]_{1.26}}{erg s^{-1}pc^{-2}}-41.17 \pm 0.9
\end{equation}

The above equation is a power-law with a slope of nearly 1.0, which represents a linear relationship.  The error is calculated from finding the slope and intercept for each individual galaxy, and finding the standard deviation of that spread.  Therefore the error represents the variations between individual galactic fits.    

The main uncertainty in the SNrate calculation is caused by the observational errors in Br$\gamma$, [FeII] and K band continuum.  The calibration uncertainty dominates the error, this is 10\% of the flux for all cases except NGC 1792, where the uncertainty is dominated by noise and is estimated to be 20\% of the flux.  If we alter the Br$\gamma$ flux by $\pm 10\%$, this translates to a change in SNrate of $10\%$.  In addition, a major systematic source of error is the choice of burst model.  Since our calculation of SNrate depends heavily on SB99, we are limited to only two star formation models, instantaneous and continuous.  In the next section, the dependency on the burst model is discussed in detail.

\section{Comparison to Continuous Star Fromation and Radio SNrate}
\label{sec:radio}
\begin{figure}[h]
\begin{minipage}{3in}
\includegraphics[width=3in]{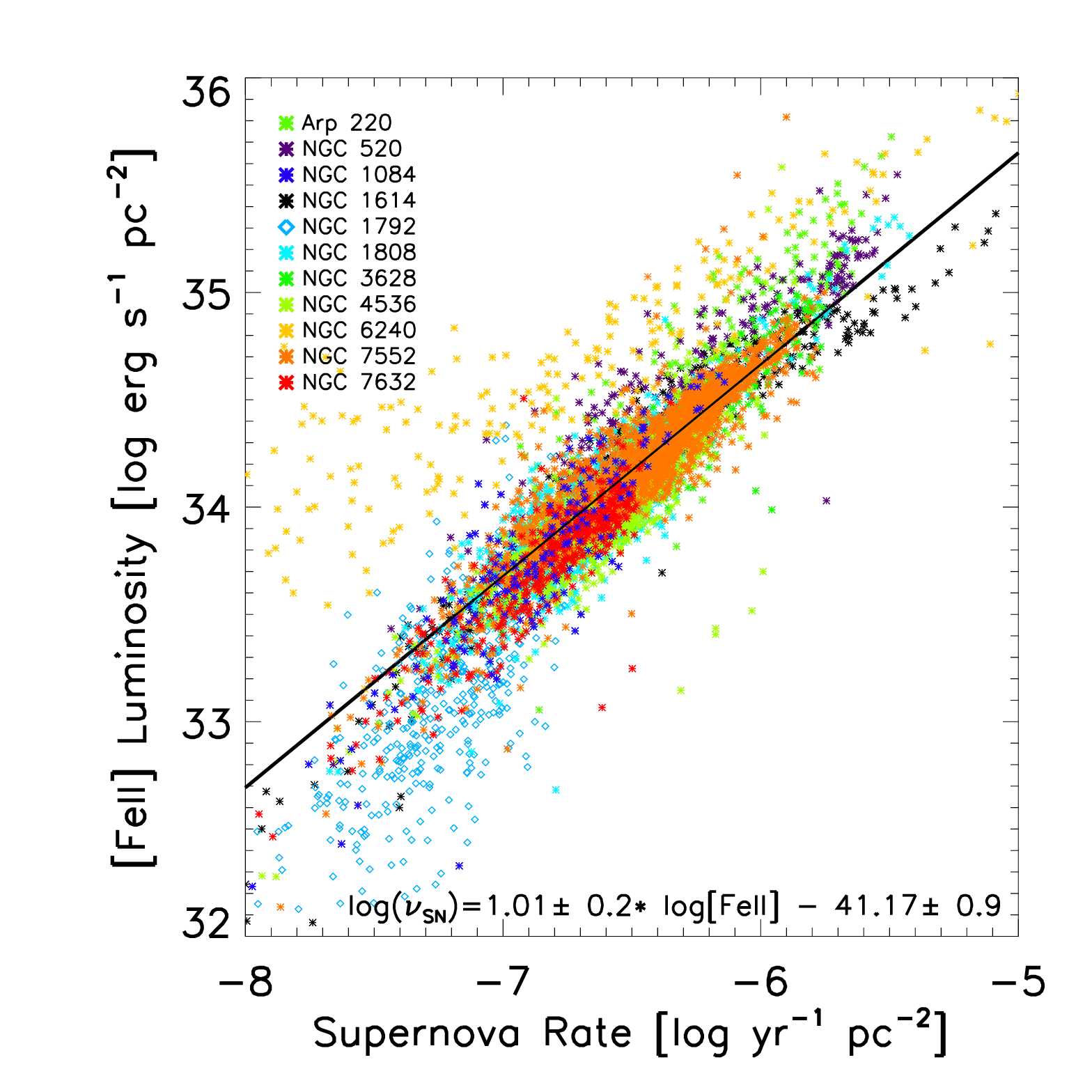}
\caption{\label{fig:pixpix}A pixel-pixel plot of SNrate, as derived from SB99, compared to [FeII] luminosity.  Each galaxy is represented with a different color and the values are normalized to 1 square parsec.  The black line represents the best fit powerlaw excluding NGC 1792.}
\end{minipage}\hspace{.5in}%
\begin{minipage}{3in}
\includegraphics[width=3in]{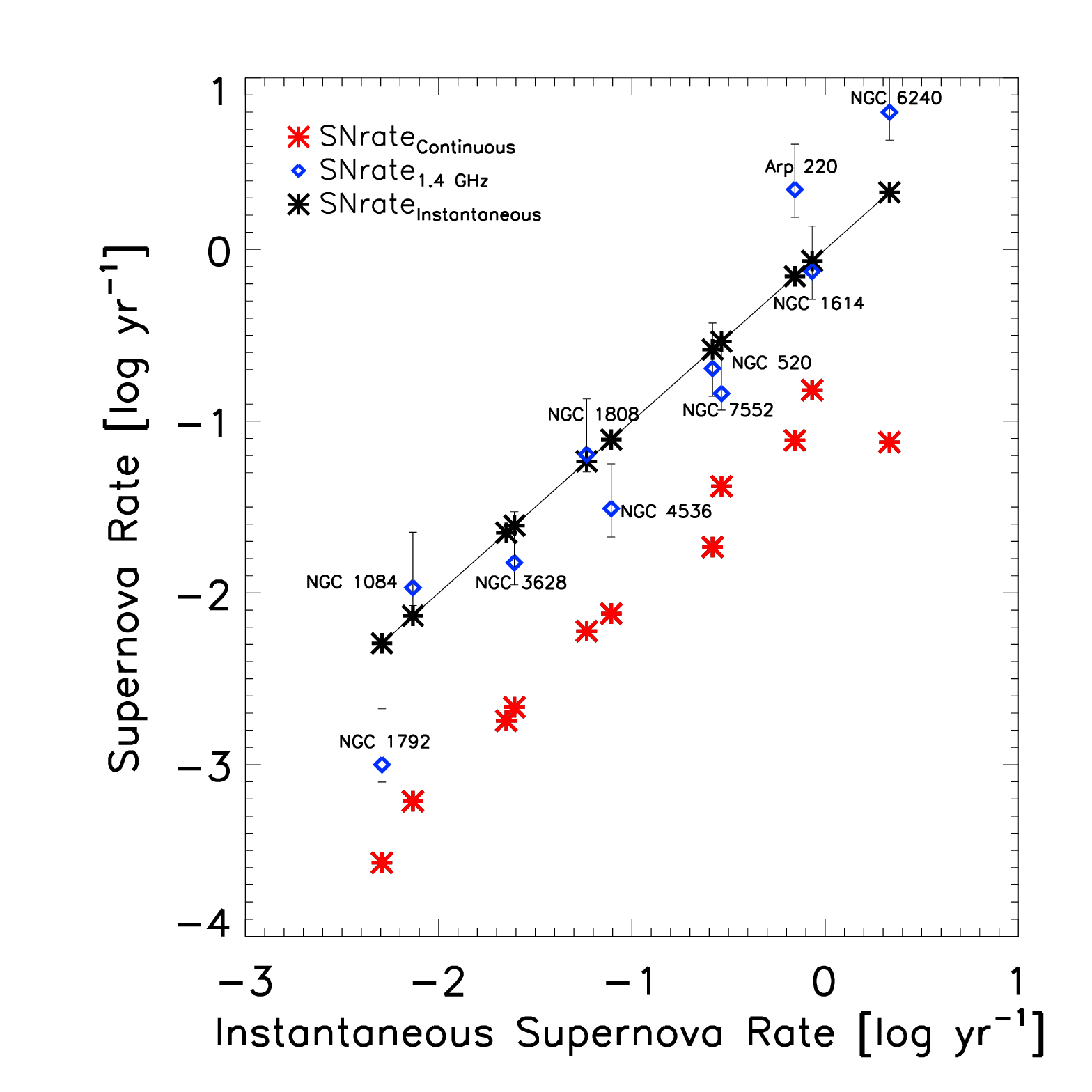}
\caption{\label{fig:radio1}Comparison of the SB99 instantaneous SNrate (x-axis) to the radio SNrate (blue diamond), the SB99 continuous SNrate (red asterisk) and for comparison the SB99 instantaneous SNrate (black asterisk).}
\end{minipage} 
\end{figure}

An initial assumption used to perform this analysis was that our galaxies are best represented by an instantaneous burst of star formation.  The other extreme case, also modeled by SB99, is the continuous star formation scenario.  It is unlikely that our galaxies are represented by either of these extreme cases, but rather somewhere in between a starburst that is extended in time.  Since radio continuum is the classical tracer of SNrate, the radio flux densities provide an independent measurement of the accuracy of the SB99 derived SNrate.  To estimate the radio SNrate, the equation given by \cite{1994ApJ...424..114H}, relating non-thermal radio luminosity and SNrate, is used along with VLA observations.  The integration area is matched to that of SINFONI.  Figure~\ref{fig:radio1} compares the SB99 instantaneous SNrate to both the radio SNrate and the SB99 continuous SNrate.  

For all galaxies, except NGC 1792, NGC 4536, NGC 6240 and Arp 220, the
radio SNrate closely matches those given by the SB99
instantaneous-burst model.  As noted before, the NIR-emission line-maps
of NGC 1792 are too noisy to be reliable.  However, NGC 4536 is a
nearly face-on galaxy with relatively low extinction and high
signal-to-noise spectra, and the NIR data should be reliable. Thus,
NGC 4536 is undergoing star formation that is closer to instantaneous but extended in time.  Arp 220 and NGC 6240 have radio
SNrates even higher than the SB99 instantaneous SNrates.  Also, in Figure~\ref{fig:pixpix}, the points for these galaxies lie above the best fit line, demonstrating a [FeII] excess.  This excess
is caused by merger-related shocks that are exciting the [FeII] in addition to the SNRs.  These powerful merger (U)LIRGs may demonstrate a limit to the
validity of the direct relationship between SNrate and [FeII] luminosity.  In addition, determining an accurate extinction in (U)LIRGs presents a challenge.  The under or over estimation of extinction leads to inaccurate [FeII] luminosities as well as Br$\gamma$ luminosities used to scale the SNrate.  However, this can only be established
by studying a larger sample of (U)LIRGs.  In any case, it appears that
the majority of (modest) starburst galaxies is well-represented by
the assumption of a (nearly) instantaneous burst of star formation.
In addition, the very good agreement between SNrates derived from the
radio continuum, and from NIR data provides added confidence in the
diagnostic strength of [FeII] as a tracer of SNrate.

\section{Conclusion}
\label{sec:conclusion}
Using SINFONI observations of 11 near-by galaxies, we performed a pixel-by-pixel analysis of the correlation between [FeII]$_{1.26}$ emission and SNrate.  First, Br$\gamma$, Pa$\beta$, [FeII]$_{1.26}$, [FeII]$_{1.64}$, and H$_{2,2.12}$ line fluxes were remeasured.  The Br$\gamma$ line flux along with the K band continuum was used to find the equivalent width of Br$\gamma$, which was then used as input into the SB99 model to estimate the age of each pixel.  From age, SB99 provides a SNrate, normalized to an initial mass of $10^6 M_{\odot}$.  The normalized SNrate is scaled by Br$\gamma$ luminosity to derive the true SNrate in each pixel.

Comparison of the [FeII]$_{1.26}$ luminosity to the SNrate (derived only from Br$\gamma$ equivalent width and luminosity) reveals a nearly linear correlation.  This relationship is valid both on a pixel-pixel basis and for the
integrated galaxy.  For the integrated [FeII] luminosity and SNrate,
the fit is remarkably tight with very little spread.  However, to
correctly determine the absolute SNrates, it is still critical to
determine whether star formation has occurred in a (nearly)
instantaneous burst, or has proceeded in a continuous fashion.
SNrates derived from radio continuum observations may be used to
distinguish these scenarios.  Most of the modest starburst galaxies
in our sample are best fitted assuming instantaneous star formation.  However, we find that the relationship breaks down for (U)LIRGs in our sample.

The use of [FeII] as a robust tracer of SNrate is a useful diagnostic tool.  Specifically, it would allow measurements of SNrates from NIR observations of distant galaxies, where individual SNR can not be resolved.  With the strong correlation found in this sample of galaxies, we can confidently say that [FeII]$_{1.26}$ is quantitatively correlated to SNrate and can be used to derive SNrate on either a pixel-pixel or integrated galactic basis.

\bibliographystyle{iopart-num}
\bibliography{proc.bib}

\end{document}